\documentclass{aastex}
\usepackage{spr-astr-addons}
\usepackage{url}

\RequirePackage{color}

\newcommand{\emaila}{wilhelm@mps.mpg.de}
\newcommand{\emailb}{bholadwivedi@gmail.com}

\def\deg{\hbox{$^{\rm o}$}}

\renewcommand{\vec}[1]{\mbox{\boldmath $#1$}}
\newcommand{\m}{{\ \mathrm m} }
\newcommand{\J}{{\ \mathrm J} }
\newcommand{\s}{{\ \mathrm s} }

\begin{document}

\title{Gravitational matter-antimatter impact interactions}

\shorttitle{Matter-antimatter impact interactions}
\shortauthors{K. Wilhelm and B.N. Dwivedi}

\author{Klaus Wilhelm}
\affil{Max-Planck-Institut f\"ur Son\-nen\-sy\-stem\-for\-schung
(MPS), 37077 G\"ottingen, Germany\\ \emaila}
\and
\author{Bhola N. Dwivedi}
\affil{Rajiv Gandhi Institute of Petroleum Technology (RGIPT), Jais,
Amethi-229304, India \\ \emailb}

\today

\vspace{1cm}

\begin{abstract}
The production of antihydrogen by several research groups provides
the opportunity to measure the gravitational behaviour of antimatter
in the gravitational field of the Earth. The predictions in the literature
range from normal attraction to repulsion. Applying our gravitational
impact model, which is of a purely phenomenological nature,
we conclude that there will be neither attraction nor repulsion under
the assumption of a symmetric antigraviton distribution near the
antihydrogen atom. However, a very small asymmetry must be expected
and could effect the conclusion.
The model, in addition, predicts normal gravitation between
antimatter and antimatter particles at large distances, but strong repulsion
at close range for matter as well as for antimatter pairs,
whereas strong attraction will result for matter-antimatter encounters.
We have further refined the model assumptions in light of recent CERN
ALPHA-g measurements that indicate a certain attraction of antihydrogen
by the gravitational field of the Earth.
\end{abstract}


\section{Introduction}  
\label{s:introd}
\subsection{Matter and hypothetical antimatter}  
\label{ss:1_1}
The physical processes controlling baryonic matter and photons in our
Solar System environment have been identified by famous
scientists, such as
Galileo Galilei,
Isaac Newton,
James Clerk Maxwell,
Roland E\"otv\"os,
Max Planck,
Hendrik Lorentz,
Albert Einstein,
Satyendranath Bose,
Enrico Fermi,
Steven Weinberg,
and many others.

In particular, the ``Standard Model'' is very successful
in formulating most aspects of particle physics, cf., e.g.
https://home.cern/science/physics/standard-model
(last accessed September 5, 2023).
Although \citet{Bor22}
in a recent paper agree with this statement,
they add in the abstract:

\begin{quote}
``The standard model of particle physics is both incredibly successful and
glaringly incomplete. Among the questions left open is the striking imbalance
of matter and antimatter in the observable universe$^1$, [...].''
\end{quote}

Note~1 refers to \citet{DinKus}:
\begin{quote}
``The origin of the matter-antimatter asymmetry is one
of the great questions in cosmology.''
\end{quote}

This asymmetry is also considered as severe problem by other authors,
for instance, by \citet{Mor58}:

\begin{quote}
``Matter made of particles, protons, electrons, and neutrons, is all about,
but anti-matter, made of antiparticles , is nowhere to be found. It is none
the less possible to manufacture it, but only at great expense.''
\end{quote}

\citet{Sch98} asked already in 1898 in a ``Holiday Dream'':

\begin{quote}
``We know positive and negative electricity, north and south magnetism, and
why not extra terrestrial matter related to terrestrial matter [...],
gravitating towards its own kind, but driven away from substances of which
the solar system is composed. [...]. The fact that we are not acquainted
which such matter does not prove its non-existence; [...].'
\end{quote}

He called this ``extra terrestrial matter'' anti-matter.\\

\citet{Dir28a} discussed a hypothetical positive electron in 1928.

\subsection{Discovery of antimatter}   
\label{ss:1_2}

The ``positive electron'' with a mass comparable to an electron was
detected 1932 in cosmic ray showers
by \citet{And32,And34}.
It was the first experimental proof that antimatter exists.

In his Nobel Lecture 1933 Paul Dirac elaborated on the positron
concept\footnote{In a paper by \citet{And33}, the editor of Physical Review
suggested the name ``positron'' for the positive electron.}:

\begin{quote}
``[...], \emph {any unoccupied negative-energy state, being a departure
from uniformity, is observable and is just a
positron.}\footnote {Emphasis by Dirac}
An unoccupied negative-energy state, or hole, as we may call it for brevity,
will have a positive energy, since it is a place where there is a shortage of
negative energy. A hole is, in fact, just like an ordinary particle, and its
identification with the positron seems the most reasonable way of getting
over the difficulty of the appearance of negative energies in our
equations. [...]. \newline
If we accept the view of complete symmetry between positive and negative
electric charge so far as concerns the fundamental laws of Nature, we
must regard it rather as an accident that the Earth (and presumably the
whole solar system), contains a preponderance of negative electrons and
positive protons. It is quite possible that for some of the stars it is the other
way about, these stars being built up mainly of positrons and negative
protons. In fact, there may be half the stars of each kind. The two kinds of stars
would both show exactly the same spectra, and there would be no way
of distinguishing them by present astronomical methods.''
\end{quote}

Not only Dirac assumed that ``negative protons and positrons'' would emit
the same spectra in stars as atoms in normal stars,
but also \citet{AlfKle} and \citet{Vla65}. \citet{Alf65} postulated that:

\begin{quote}
``Such antiatoms should have the same
properties as ordinary atoms. For example they could
build up chemical compounds similar to ordinary
chemical compounds, and they should emit spectral
lines of exactly the same wavelengths as ordinary
atoms.''
\end{quote}

\subsection{Production of antimatter}   
\label{ss:1_3}

Modern experiments can now produce antimatter and have confirmed with high
accuracy that antihydrogen emits indeed the same spectral lines as hydrogen
in our Earth environment
\citep{Paretal11,Amoetal12,Ahmetal17,Ahmetal20}.
These results imply that hydrogen would emit the same spectra
in an antimatter environment and, of course, antihydrogen as well.

The important progress achieved, in particular, by the CERN Collaborations

\begin{enumerate}
\item
ALPHA (Antihydrogen Laser Physics Apparatus)
\cite[cf., e.g.][]{Amoetal14,Ber18,Ahmetal18a,Ahmetal18b},
\item
AEgIS (Antimatter Experiment: gravity, Interferometry, Spectroscopy)
\cite[cf., e.g.][]{Keletal,Tesetal15,Bruetal17,Amsetal21}, and
\item
GBAR (Gravitational Behaviour of Antihydrogen at Rest)
\cite[cf., e.g.][]{Peretal15,Man19,CriKol}
\end{enumerate}

in producing antihydrogen can now also be applied to studies of the
gravitational behaviour of matter-antimatter
systems.

\citet{NieGol} considered many arguments against ``Antigravity'' and
discussed, in particular, arguments of \citet{Mor58},
\citet{Sch59}, and \citet{Goo61}. In the
conclusion section, they mention that a consensus exists
that an antiproton gravity experiment is important and further point out:

\begin{quote}
``Somewhat paradoxically, it turns out that the more precisely anomalous
gravitational effects are
ruled out in earth-based matter-matter experiments,
the more unrestricted is the possibility that there
can be a significant anomalous gravitational acceleration of antimatter.''
\end{quote}

One of the arguments taken from \citet{Mor58} assumes that a photon
is blue-shifted when it falls in a gravitational field.
\citet{Oku00} made the following statements in Chapter~6 of
``Photons and static gravity'':
The frequency and energy of a photon in a static gravitational field
are constant, however, the momentum and the
wavelength change, see also \citet{Okuetal,WilDwi19}.
Morrison assumed that a red shifted photon has
insufficient energy. The confusion is that red shift can refer to wavelength
or frequency. Morrison obviously relates the shift to frequency, which is
wrong, and it is thus questionable,
whether the statement that energy is not conserved by the ``antigravity''
theory can be justified by this assumption.

\citet{CalDva}
expect that the experiments will show that antimatter is attracted by matter
as the interacting field in Einstein's theory has a spin = 2, cf., e.g.
\citet{OgiPol}. Particle and antiparticle attraction is also
predicted by \citet{Sch59}.

Arguments for matter-antimatter repulsion are presented, among others, by
\citet{Cab10},
\citet{Vil11,Vil13,Vil15}, \citet{Haj20a}, \citet{HajWal}, and
\citet{Chaetal21}. \citet{Sch79} considered ``antigravity'' as well as
``antigravitons''.\footnote{The name ``Graviton'' has probably been coined
1934 by D.I.~Blokhintsev \& F.M.~Gal'perin, cf. \citet{Rov00}.}

No experimental violation of the
Charge, Parity, and Time theorem (CPT)\footnote{A proof of this theorem has been
given by G. \citet{Lue57}.} has been observed
in the matter-dominated Universe, cf. \citet{NieGol}, who also state that the
CPT theorem does not determine the attraction or repulsion between matter
and antimatter. Therefore, a violation of the principle of equivalence
is not excluded by CPT. The invariance of the CPT theorem is now also
experimentally tested with antibaryons \citep{Ahmetal17,Bor22}.

The Weak Equivalence Principle (WEP) \citep{Ein11,Ein16} has been experimentally
confirmed many times with matter particles
\cite[cf., e.g.][]{Eoeetal,Goo61,NieGol,Touetal19}. The planned
experiments with antihydrogen will be of great importance to show,
whether or not WEP is valid for antimatter in a matter
environment \citep{Amsetal21,Chaetal21}.

Another question in relation to matter and antimatter is whether the
observed asymmetry is present in the complete Universe or if
this is only a local aspect. In line with Dirac's suggestion, proposals are
presented that assume a symmetric creation of the Universe. They are discussed,
for instance, by \citet{Gol56,Sak66,Klein01,Vil13,WillJung}, and
\citet{Haj20b}.
The separation
of the matter and antimatter could be achieved, when small density
inhomogeneities are enlarged by gravitational instabilities.
\citet{Gol56} speculated that the statistical fluctuations might have
produced galaxies and ``antigalaxies''.

\citet{Haj20a} summarized three models with repulsion between matter and
antimatter. Matter attracts, of course, matter. The first two models, called
``wild'' by Hajdukovic, \newline
``assume a symmetric Universe with equal amounts of matter and antimatter'':
\begin{enumerate}
\item
The Dirac-Milne Cosmology, cf., e.g. \citet{BenCha}, with
gravitational repulsion both between matter and antimatter as well as
between antimatter and antimatter.
This leads to a CPT violation.
\item
The Lattice Universe, cf., e.g. \citet{Vil13}, with antimatter-antimatter
attraction. There is no CPT violation.
\item
Quantum vacuum fluctuations and virtual gravitational
dipoles (GD) \citep{Haj13,Haj20b,HajWal}.
Again there is attraction between antimatter and
antimatter and no CPT violation.
The Universe might alternate between matter and antimatter cycles.
\end{enumerate}

\citet{BanKro} conclude, however, that the gravitational dipole (GD) model
does not satisfy galaxy rotation curve observations
and Solar System constraints, in agreement with \citet{Ior19}.

Many more papers discuss the antimatter experiments and the results achieved or
expected. We could only mention a small selection of them as an introduction
to processes of matter-antimatter interactions
based on our proposed gravitational impact model
\citep{WilWilDwi,WilDwi20}.

\section{Gravitational impact model}  
\label{s:Results}

\subsection{The model}  
\label{ss:2_1}

Our heuristic model is based on the idea of a gravitational impact concept
presented by \citet{Fat90} at the end of the seventeenth century.
Small corpuscles moving with high speed interact with masses.
The attraction results from a shielding effect caused by the bodies.
Many objections
have been raised against the shielding concept, in particular,
the disturbing effect of a third body between two masses was discussed by
\citet{Dru97}. We felt that this problem can be overcome by replacing
the shielding by an interaction of hypothetical massless particles
with the masses.
Quadrupoles\footnote{Quadrupoles with two positive and two negative
elementary charges
have been proposed.} are good candidates, because they have small
interaction energies with positive and negative electric charges and,
in addition, can be constructed with a spin of $S = \pm 2$ \citep{WilWilDwi}.
We will call the quadrupoles now {\em gravitons}, cf. Footnote~3.
They have no mass and a speed of light,
the mean value of their energy is $T_{\rm G}$ and
that of the magnitude of the  momentum $|\vec{p}_{\rm G}|$.
Estimates on these quantities will be given in Subsection~\ref{ss:3_4},
but the spectral distribution is unknown.

\begin{figure}
\includegraphics[width=\columnwidth]{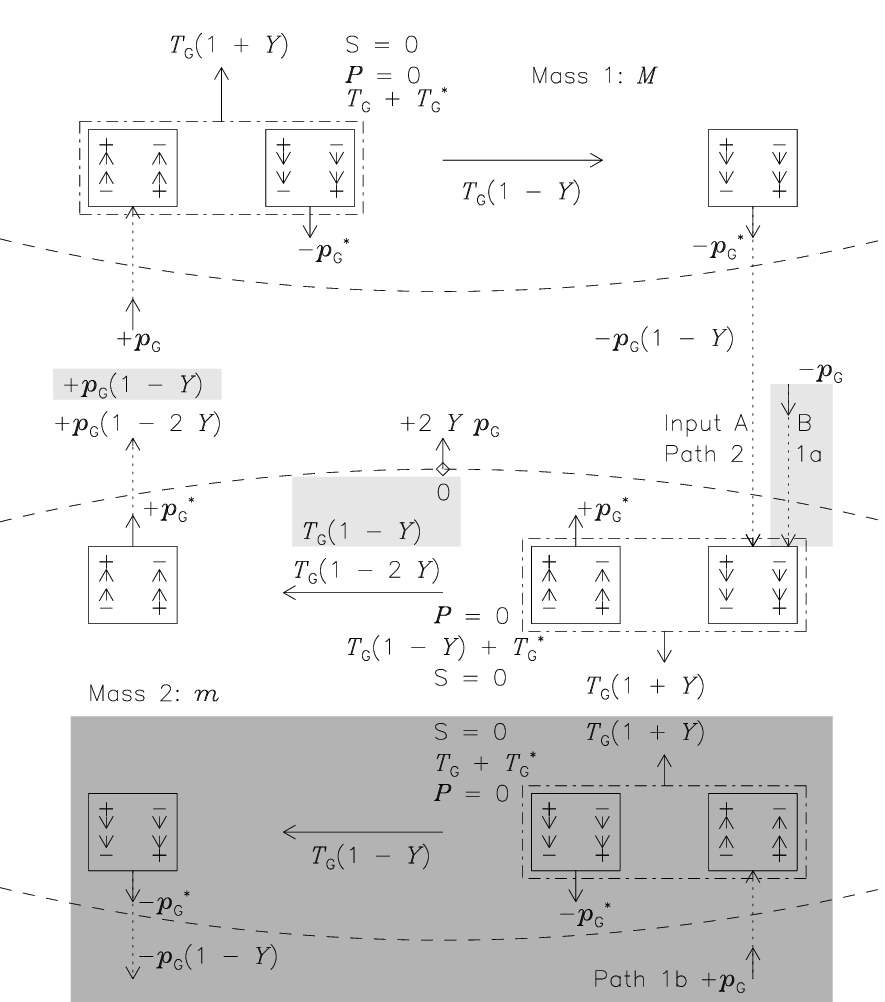}
\caption[Matter-matter gravitation]{Matter-matter gravitation
in a background flux of gravitons with momentum
$\pm\vec{p}_{\rm G}$ and spin (shown by arrows in the square boxes,
which indicate the gravitons)
in the same direction. The virtual gravitons are
denoted by $\vec{p}^*_{\rm G}$. Their $\pm$ signs indicate directions up or
down
on the diagram.
Without interaction between Mass~1 (of mass~$M$) and Mass~2 (of mass~$m$),
both masses will be at rest, because the interaction processes are symmetric
in all directions.
This is shown for $m$ where the action introduced by
$-\vec{p}_{\rm G}$ in Input~B and Path~1a
(light shading) will be compensated by the inverse action
of $+\vec{p}_{\rm G}$ on Path~1b from below (dark shading).
If, however, the reduced
graviton $-\vec{p}_{\rm G}(1-Y)$ from
Mass~1 in Input~A replaces $-\vec{p}_{\rm G}$ from the background graviton
distribution, then the momentum balance for Mass~2 becomes
$+ 2\,Y \vec{p}_{\rm G}$, see Equation~(\ref{eqn:matter_matter}).
The exact energy reduction
$T_{\rm G}(1 - Y)^2$ on Path~2 has been approximated by $T_{\rm G}(1 - 2\,Y)$
in view of the very small value of $Y$.
The effect of Mass~2 on Mass~1
has not been determined in this diagram, but is in line with the momentum
conservation principle as shown in \citet{WilDwi20}.
By reversing all matter into antimatter and gravitons into
antigravitons, we would obtain the same results.\\
The graviton background will be supplemented by antigravitons
in Subsection~\ref{ss:2_3}. Since their
interactions are not balanced, a small correction must be
taken into account, see Subsection~\ref{ss:3_2}.}
\label{fig:M_g_m}
\end{figure}

It was, however, necessary to introduce
energy and momentum losses of the gravitons during the
interaction with matter, i.e.
$T_{\rm G}$ will be reduced to $T_{\rm G}(1 - Y)$, when it is re-emitted, and
$|\vec{p}_{\rm G}|$ to $|\vec{p}_{\rm G}(1 - Y)|$, where $0 < Y \ll 1$.
The attraction between matter particles is accomplished by the exchange
of reduced-momentum gravitons between them that do not balance the
impact of the gravitons from the opposite direction.
The gravitational impact interaction process
is demonstrated in Figure~\ref{fig:M_g_m} with only one of very many reduced
gravitons in Input~A (Path~2). In a sense, the
particles will not attract each other, but are pushed together, in
close analogy to Fatio's process.

The long-range gravitational force is thus described by a
heuristic impact model
with hypothetical massless entities propagating at
the speed of light in vacuum and transferring momentum and
energy between massive bodies through interactions on a local basis.
In the original publication \citep{WilWilDwi},
a spherically symmetric emission of secondary entities have
been postulated. The potential energy problem in a gravitationally
bound two-body system with arbitrary mass distribution
has been studied in \citet{WilDwi15}
in the framework of the impact model of gravity.
This study has indicated that an anti-parallel emission of a secondary
quadrupole\,--\,now called graviton\,--\,with
respect to the incoming one
is more appropriate, because it does not violate the energy conservation
principle. It could be shown that in the latter case the difference of
the potential gravitational energy during an approach of the masses
corresponds to the difference of the sum of the reduced energy of the
gravitons on their way between the masses.

The model could successfully be applied to secular perihelion advances
\citep{WilDwi14} and radial accelerations of disc
galaxies \citep{WiDwGala,WiDwGalb}. In analogy to the graviton interaction,
we also formulated a heuristic model of the electrostatic forces.
Repulsion of particles with
the same charge and attraction of oppositely charged particles
could be achieved with dipoles interacting directly or
indirectly\footnote{Direct and indirect interactions are defined
in the case of gravitons and antigravitons
in Subsections \ref{ss:2_2} and \ref{ss:2_3}, respectively,
in analogy to the electrostatic situation.} with the
particles \citep{WilDwi20}.

\subsection{Matter and matter interactions}
\label{ss:2_2}
The interactions
of the gravitons~$\pm\vec{p}_{\rm G}$ with masses $M$ and $m$, respectively,
occur with virtual gravitons (emitted from a particle with
$\pm{\vec{p}}^*_{\rm G}$,) and result in zero spin and momentum
(assuming $|\vec{p}_{\rm G}| = |\vec{p}^*_{\rm G}|$) with
an energy of $2\,T_{\rm G}$ \citep{WilDwi20}. The reduced energy
$T_{\rm G}(1 - Y)$ will liberate a virtual graviton.
This process will be called a \emph{direct interaction}. It is
demonstrated in Figure~\ref{fig:M_g_m} on top of Mass~2 by light shading
(Input~B, Path~1a) and below by dark shading (Path~1b).
The momentum balance is zero in this case. If, however, a graviton with
reduced momentum arrives from Mass~1 on Input A the balance is
%
\begin{eqnarray}
-\vec{p}_{\rm G}(1-Y)-\vec{p}_{\rm G}(1-2\,Y)+\vec{p}_{\rm G}
+\vec{p}_{\rm G}(1-Y) = \nonumber \\
+ 2\,Y \vec{p}_{\rm G}~~,
\label{eqn:matter_matter}
\end{eqnarray}
for only one of the many reduced gravitons required.
We assume that the total number of reduced gravitons
per time interval, e.g. $1~\s$, is $n$ for $1~g = 9.81\,\m\,\s^{-2}$
gravitational acceleration of $m$ towards the Earth with
mass~$M$. This gives a momentum balance
of $+ 2\,n\,Y \vec{p}_{\rm G}$.

Gravitational processes
with antimatter will require the addition of antigravitons in the next
subsections.

It is assumed that the particles Mass~1 and Mass~2 are far apart relative
to their sizes and thus
the double-reduced graviton on the left side is not expected to interact
with Mass~1.
We consider subatomic particles here.
Large solid bodies and the situation, when
the particles approach each other, are discussed in Subsections~\ref{ss:2_4}
and \ref{ss:2_5}.

\begin{figure}
\includegraphics[width=\columnwidth]{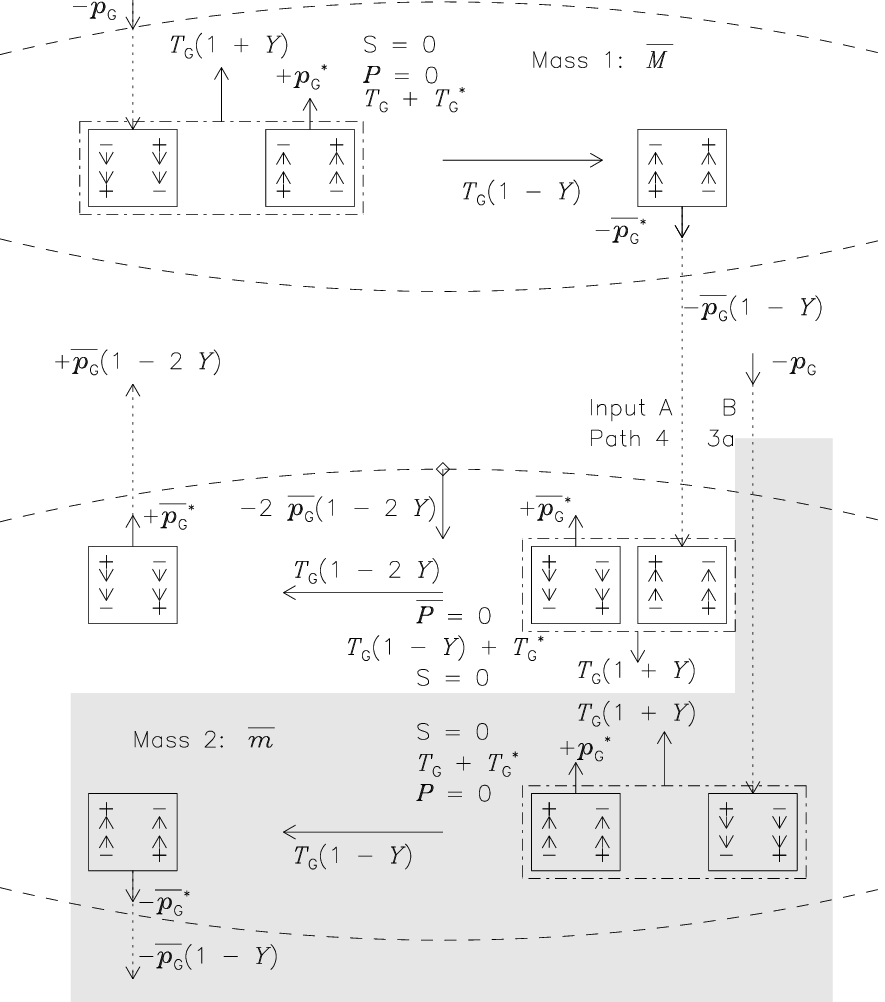}
\caption[Antimatter-antimatter interaction]
{Antimatter-antimatter interaction in a
graviton environment.
Both Masses 1 and 2 consist of antimatter ($\bar{M}$ and $\bar{m}$).
If there is no interaction between them, they are
in equilibrium, cf. Path~3a (shaded area) and a corresponding
Path~3b as shown in Figure~\ref{fig:M_g_am}.
If a reduced antigraviton $-\bar{\vec{p}}_{\rm G}(1 - Y)$
from Mass~1 is interacting with Mass~2 (Input~A)\,--\,eliminating the process
on Path~3a that will be called indirect interaction\,--\,the
momentum balance for Mass~2 then gives
in Equation~(\ref{eqn:antimatter_antimatter_1}) a momentum of
$-2\,\bar{\vec{p}}_{\rm G}(1-2\,Y)$.
Mass~2 is strongly repulsed. For reasons of symmetry the same is true for
Mass~1. By reversing all antimatter into matter and gravitons into
antigravitons, we would obtain the same result.\\
The same result would also be obtained for a matter-matter interaction
in an antigraviton environment.\\
If, however, an ambigraviton
environment according to Subsection~\ref{ss:2_3} is assumed,
we get the configuration of Figure~\ref{fig:M_g_m}, but for antimatter.}
\label{fig:AM_g_am}
\end{figure}
%
\begin{figure}
\includegraphics[width=\columnwidth]{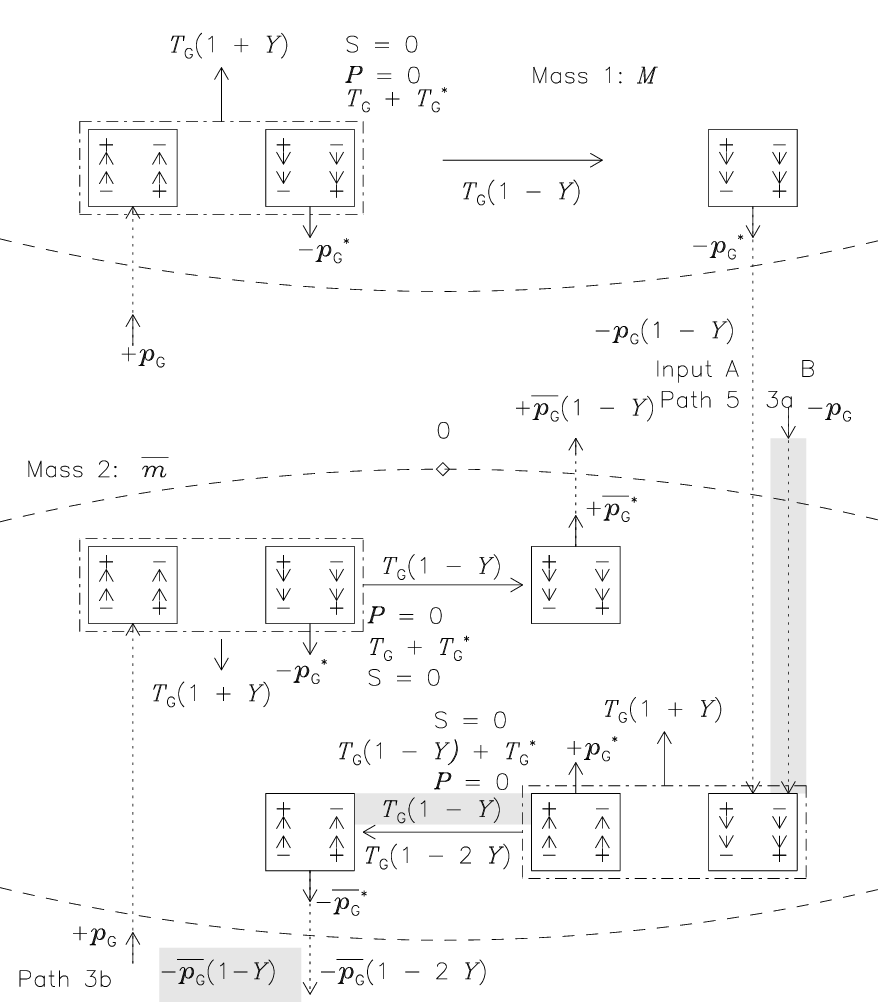}
\caption[Matter-antimatter gravitation]
{Matter-antimatter gravitation in graviton environment.
With no interaction between $M$ and $\bar{m}$, the
graviton $-\vec{p}_{\rm G}$ from the background
in Input~B (light shading) would be
compensated by $+\vec{p}_{\rm G}$ from below. Mass~2 would be at rest.
The reduced graviton $-\vec{p}_{\rm G}(1-Y)$
arriving from Mass~1 at Mass~2 (Input~A), which replaces
$-\vec{p}_{\rm G}$, cannot combine with $+\bar{\vec{p}}^*_{\rm G}$,
because a spin~$S = 4$ would result.
The other side of Mass~2 would provide an appropriate
partner in the destruction phase for an indirect interaction.
The momentum balance for Mass~2 then is zero,
see Equation~(\ref{eqn:matter_antimatter}).
Mass~2 would be at rest in this case. \\
An ambigraviton environment
with symmetric direct interactions of antigravitons
would not change the situation.
Note, however, that the discussion in Subsection~\ref{ss:3_2}
indicates a small asymmetry.}
\label{fig:M_g_am}
\end{figure}

\subsection{Antimatter and antimatter interactions}   
\label{ss:2_3}
The impact model has up to now only be applied to matter particles and
photons. In Figure~\ref{fig:AM_g_am}
the situation is shown for two antimatter
bodies in the same graviton environment as in Figure~\ref{fig:M_g_m}.
Antimatter is assumed to emit gravitons with opposite spin directions
in their creation phase as shown in the lower part of Figure~\ref{fig:AM_g_am}
for Mass~2. They will be called virtual \emph{antigravitons}.
In line with this definition, real antigravitons
have opposite directions of spin and momentum in contrast to gravitons.
The shaded area illustrates an \emph{indirect interaction} caused by the fact
that the interaction of $-\vec{p}_{\rm G}$ with $+\bar{\vec{p}}^*_{\rm G}$
in the creation phase
would give a forbidden spin of $S = 4$, whereas the annihilation
phase (at the bottom of Mass~2) reverses the spin direction and a total spin
of zero will result.
Since the indirect process is much more complicated than
the direct one, we expect only a very small fraction of indirect interactions
compared with direct ones. Nevertheless, these indirect interaction affect
the symmetry of the graviton and antigraviton distributions and will be
discussed in detail in Subsection~\ref{ss:3_2}.
\begin{figure}
\includegraphics[width=\columnwidth]{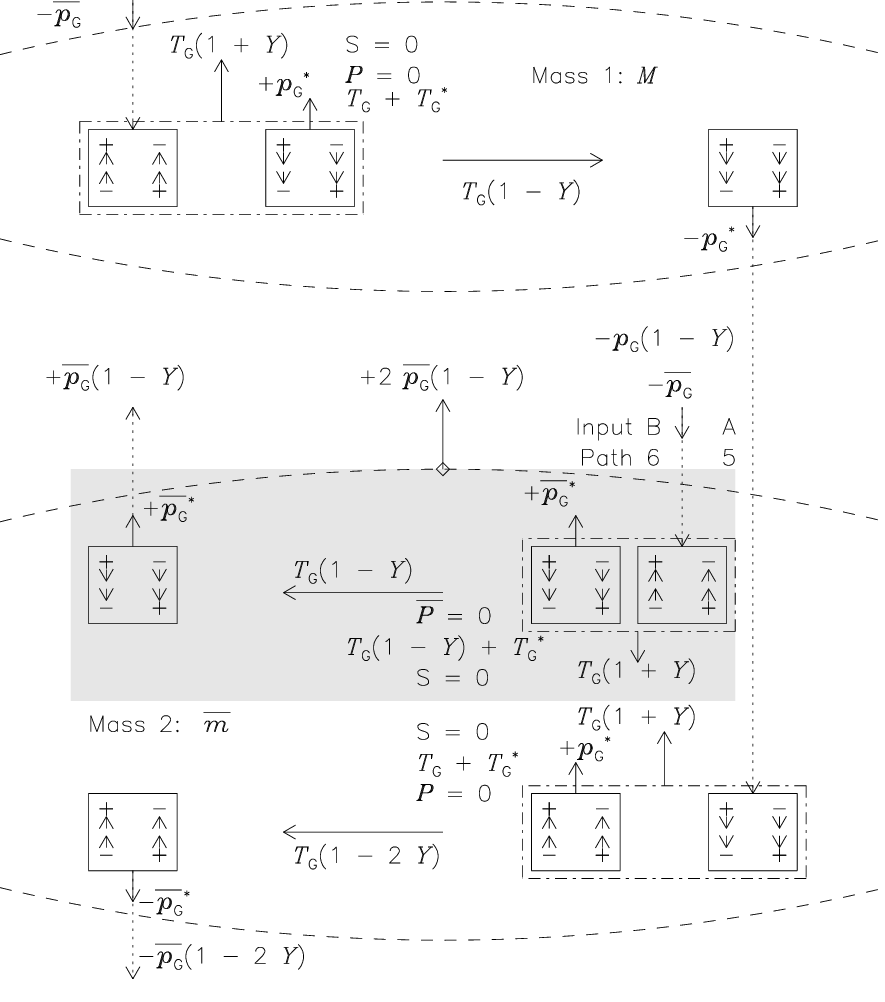}
\caption[Matter-antimatter antigravitation environment]
{Matter-antimatter gravitation in antigraviton environment.
If there is no action from Mass~1 on
Mass~2 (shaded area; Input~B), then Mass~2 is
in equilibrium. The interaction process on the opposite side, in analogy to
the dark shaded area in Fig.~\ref{fig:M_g_m}, is not shown here for antimatter
and antigravitons. It will result in $+\bar{\vec{p}}_{\rm G}$
and $+\bar{\vec{p}}_{\rm G}(1-Y)$.
If a reduced graviton $-\vec{p}_{\rm G}(1-Y)$ from Mass~1
is active (Input~A)\,--\,eliminating the
process in the shaded
area of Path~6\,--\,the momentum balance for Mass~2 is
$+2\,\bar{\vec{p}}_{\rm G}(1 - Y)$,
see Equation~(\ref{eqn:antigraviton}).
It is attracted by Mass~1.
The inclusion of gravitons, however,
would give a situation as in Figure~\ref{fig:M_g_am}.}
\label{fig:M_ag_am}
\end{figure}
If Input~B is replaced by Input~A,
the evaluation of the momentum balance is
%
\begin{eqnarray}
-\bar{\vec{p}}_{\rm G}(1 - Y)-\bar{\vec{p}}_{\rm G}(1 - 2\,Y)
+\vec{p}_{\rm G}-\bar{\vec{p}}_{\rm G}(1-Y) = \nonumber \\
-2\,\bar{\vec{p}}_{\rm G}(1-2\,Y)~~,
\label{eqn:antimatter_antimatter_1}
\end{eqnarray}
assuming
$|\pm\vec{p}_{\rm G}| = |\pm\bar{\vec{p}}_{\rm G}|$.

Considering the very small value of
$Y \le 10^{-20}$, cf. Figure~3 of \citet{WilDwi20}, the resulting repulsion would be
many orders of magnitude larger than the attraction implied by
Equation~(\ref{eqn:matter_matter}). This might, however, be in conflict
with observations referenced in Subsection~\ref{ss:1_3}
that spectral lines of antihydrogen
observed on Earth are not different from corresponding hydrogen lines,
because the exchange forces between the proton and electron, respectively,
antiproton and positron, in hydrogen and antihydrogen would be very
different and could affect the electrostatic forces. Details depend on
the value of $Y$ and the uncertainties of the spectral measurements.

What appears to be a severe problem for this interaction model, will
eventually turn out to be extremely positive, because the assumption
of a pure graviton environment in both figures was completely arbitrary
and an antigraviton environment could have been added.
Therefore, we agree with the statement of \citet{Alf65}:

\begin{quote}
``It is postulated that a cosmological theory should be matter
antimatter
\newline
symmetric, and introduce no ad hoc laws of
nature.''
\end{quote}

Thus we have to realize that in both figures
a graviton-antigraviton environment would be
appropriate\\ (\emph{ambigravitons}) to modify the
expression ``ambiplasma'' \cite[cf.][]{Alf65}.
Isolated matter and antimatter particles both experience
balanced direct and indirect
interactions in such an ambigraviton environment.

In Figure~\ref{fig:AM_g_am} the Input~A then replaces Input~B
and the process on Path~4 corresponds to
the inverse version of Path~2 in Figure~\ref{fig:M_g_m}
with a momentum balance of
%
\begin{eqnarray}
-\bar{\vec{p}}_{\rm G}(1 - Y)-\bar{\vec{p}}_{\rm G}(1 - 2\,Y)
+\bar{\vec{p}}_{\rm G}+\bar{\vec{p}}_{\rm G}(1-Y)= \nonumber \\
+2\,Y\,\bar{\vec{p}}_{\rm G}~~,
\label{eqn:antimatter_antimatter_2}
\end{eqnarray}
which is the same as for matter-matter interactions.

\subsection{Matter and antimatter interactions}    
\label{ss:2_4}
The Input~B in Figure~\ref{fig:M_g_am} demonstrates on Path~3a
an indirect interaction. A graviton~$-\vec{p}_{\rm G}$
traverses $\bar{m}$ (Mass~2) and exits as antigraviton with a momentum
$-\bar{\vec{p}}_{\rm G}(1 - Y)$. A transformation from
a graviton to an antigraviton is shown, because antimatter can only emit
antigravitons in our model. The impact of this traversal
on $\bar{m}$ will be balanced by the effect of the graviton~$+\vec{p}_{\rm G}$
entering at the lower left-hand corner on Path~3b.
This figure finally shows the matter-antimatter
process as far as the gravitons are concerned by considering Input~A.
The reduced graviton indirectly interacts with Mass~2 on
Path~5 and results in a momentum balance of zero:
%
\begin{eqnarray}
-\vec{p}_{\rm G}(1-Y)+\bar{\vec{p}}_{\rm G}(1-2\,Y)+\vec{p}_{\rm G}-
\bar{\vec{p}}_{\rm G}(1-Y)=0 .
\label{eqn:matter_antimatter}
\end{eqnarray}
The addition of an \emph{undisturbed} antigraviton distribution also cannot
provide a momentum effect on antimatter, and no attraction or repulsion is
expected under these conditions. However, an indirect interaction
of an antigraviton with
Mass~1 will cause a disturbance. Arguments are
presented in the next subsection indicating that this effect is small.

From Figure~\ref{fig:M_ag_am}, we find for one reduced graviton
%
\begin{eqnarray}
-\vec{p}_{\rm G}(1-Y)+\bar{\vec{p}}_{\rm G}(1-2\,Y)+
\bar{\vec{p}}_{\rm G}+\bar{\vec{p}}_{\rm G}(1-Y) = \nonumber \\
+2\,\bar{\vec{p}}_{\rm G}(1 - Y)~~,
\label{eqn:antigraviton}
\end{eqnarray}
and a total momentum balance of $+ 2\,n\,\bar{\vec{p}}_{\rm G}(1 - Y)$,
cf. discussion in the context of Equation~({\ref{eqn:matter_matter}}).
Matter-antimatter gravitation in an antigraviton environment thus
leads to unrealistic results for distant particles.
Ambigravitons are required to avoid this situation.
Input~A in Figure~\ref{fig:M_ag_am} would then replace a graviton and not an
antigraviton of Input~B, leading to zero momentum balance as
in Figure~\ref{fig:M_g_am}. It should also be mentioned that
Figures~\ref{fig:M_g_am} and \ref{fig:M_ag_am}
imply a violation of the momentum conservation principle without
ambigravitons.
In an ambigraviton environment, the relation between Mass~$M$ and
Mass~$\bar{m}$ can be inverted without violating the momentum balance.

In discussing Figure~\ref{fig:M_g_m} in Subsection~\ref{ss:2_2},
it has been pointed out that
the interacting particles (Mass~1 and Mass~2) should be far apart.
It is necessary at this stage to clarify that the interacting
particles are not close together for an antimatter experiment in the
gravitational field of the Earth.
The emissions of virtual gravitons and
antigravitons that interact in our model with real gravitons and
antigravitons occur from subatomic particles, such as electrons,
protons and neutrons. Larger bodies are conglomerations of these particles
and are rather transparent for gravitons and antigravitons (as discussed at
the end of Subsection~\ref{ss:2_5}), before an interaction with
a subatomic particle happens. For a spherically symmetric body, which is a good
approximation for the Earth, the gravitational attraction of a particle
above the surface of the Earth is controlled by the distance to its center.
Therefore, we can assume for the antihydrogen gravity experiment
a mean separation of an Earth radius between the interacting subatomic
particles. The result of Equation~(\ref{eqn:matter_antimatter}) with
zero momentum thus is relevant for the antihydrogen experiment, assuming
no indirect interactions, which would affect the symmetry of the ambigraviton
environment near particles.

This surprising prediction was the reason for writing this article, since
in the literature either attraction or repulsion of antimatter
in the gravitational field of the Earth had been discussed.
Should our prediction in future measurements be confirmed, this would
support the model. If attraction or repulsion is observed, the model
would have to be reconsidered.

While working on a revision of this manuscript, a Nature
article was published on September 28, 2023 with ALPHA-g measurements by
\citet{Andetal23}. They found that the best fit of the local acceleration of
antimatter towards the Earth was $0.75~g$
(with relative statistical and systematic uncertainties of $\pm 18.5~\%$
and $\pm 21~\%$ as relative simulation uncertainty). The probability that
the result could occur under the assumption that the gravitational field
of the Earth does not act on antihydrogen was $2.9 \times 10^{-4}$.
The probability that repulsive gravity is consistent with the data
is less than $10^{-15}$.

In view of these findings\footnote{It should be noted that \citet{Du2023}
interpreted the ALPHA-g measurements differently and concluded that for some
antihydrogen atoms repulsive antigravity exists.}, modifications of our model assumptions will be
discussed in Subsection~\ref{ss:3_2}.

\subsection{Short range forces}   
\label{ss:2_5}

If the subatomic particles are at close range, the situation
is very different from the configurations discussed so far.
In this context, we emphasize the prediction of \citet{Fle17}
(without referring to ``The Matter-Antimatter Dipole''
hypothesis\footnote{Gravitational dipole (GD) models
are controversially discussed in the literature, e.g.
\citep{Haj10,Haj13,Haj20a,Haj20b,HajWal}.}):

\begin{quote}
``It is clear from known physics that matter and antimatter behave
in such a way that matter is repelled from matter at
short-range\footnote{Note added by KW and BND: Neutron stars could
serve as an example. We do not know, what happens in black holes, but as a
``wild'' speculation we could refer to Subsection~4.1 of \citet{WilWilDwi},
where we found that the mass of a particle might reside within the sum
of its virtual quadrupoles, and suggest that in a black hole the virtual
gravitons of all of its constituents combine.}
and by extension, antimatter is repelled
from antimatter. It is also clear that at
short-range, matter and antimatter attract.''
\end{quote}

Is our impact model in agreement with these expectations?
In Figure~\ref{fig:M_g_m} with the assumption of ambigravitons
two effects can be identified:

\begin{enumerate}
\item
If the particles approach each other, the double reduced
graviton~$+\vec{p}_{\rm G}(1-2\,Y)$ on the
left-hand side resulting from Path~2 will interact with Mass~1 and enlarge
the attraction, because it will reduce the graviton momentum
on Input~A by $2\,Y\,\vec{p}$. There will be a tendency of further
enhancements, because
the reductions at Input~A will also become more important.
\item
This cycle comes, however, soon to an end, when Mass~1
prevents some antigravitons~$-\bar{\vec{p}}_{\rm G}$ to reach Mass~2.
They would have traversed $m$ on a path corresponding to Path~3a, shown in
Figure~\ref{fig:AM_g_am} and Figure~\ref{fig:M_g_am}
for gravitons and
$\bar{m}$, compensated by the opposite Path~3b
in Figure~\ref{fig:M_g_am}. The shielding of antigravitons is accompanied
by interactions of reduced gravitons from Mass~1 with Mass~2.
This causes a very strong repulsion,
as Path~4 in Figure~\ref{fig:AM_g_am} demonstrates for antimatter
particles in a graviton environment.
\end{enumerate}

In Figure~\ref{fig:M_ag_am} finally a close encounter of matter and
antimatter can be set up by supposing that the
antigraviton of Path~6 is indeed shielded by Mass~1 and
is replaced by the reduced graviton on Path~5. This gives a very strong
attraction.
It can be concluded that the model yields the results as expected.

Nevertheless, at least two questions have to be answered:
\begin{enumerate}
\item
The indirect interaction  with an antiparticle
(see, for instance, Path~3a in Figure~\ref{fig:M_g_am}) changes a graviton
into a reduced antigraviton .
An inversion from an
negative antigraviton into a negative reduced
graviton by Mass~1 is shown in Figure~\ref{fig:M_ag_am}.
The question thus is,
whether in a matter dominated region all antigravitons will soon be converted
to gravitons\,--\,with an opposite process in an antimatter environment?
\item
The observations on Earth require that a
graviton-antigraviton environment must be present in our model
to avoid unusual
effects, cf. Subsection~\ref{ss:2_4}. Does the Earth reverse many
antigravitons?
\end{enumerate}

For both questions, it has to be realized that gravitons can travel
very long distances through solid bodies, before they interact with
a particle. This is indicated by the small effect that
multiple interactions
within the Sun have on secular perihelion advances \citep{WilDwi14} and
that only about five interactions of a graviton on its way from the center
of a disk galaxy to the boundary are expected \citep{WiDwGala,WiDwGalb}.
The expectation in Subsection~\ref{ss:2_3} that indirect interactions
are much less frequent than direct ones will be supported by the discussion
in Subsection~\ref{ss:3_2}.
Since we assume equal amounts of matter and antimatter in the Universe,
the long travel distances will on average maintain a constant
graviton-antigraviton mixture and the Earth will only convert a small
fraction of antigravitons into reduced gravitons.

\section{Performance of the impact model}   
\label{s:Perform}

The impact model could successfully be applied to many gravitational
observations in the matter environment, cf. \citet{WilDwi20} and references therein.
The experimental progress in the production of antihydrogen as outlined in
Section~\ref{s:introd} motivated us to expand the proposed model to
antimatter problems as well.

\subsection{Symmetry}    
\label{ss:3_1}

The result of Equation~(\ref{eqn:antimatter_antimatter_1}) obtained from
Figure~\ref{fig:AM_g_am} demonstrates that a graviton
environment as assumed in \citet{WilWilDwi,WilDwi20}
would lead to a very strong repulsion between antimatter and antimatter,
if gravitons interact at all with antimatter. Both alternatives
might, however, not be consistent with the observations described in
Subsection~\ref{ss:1_3} that antihydrogen emits the same spectral lines as
hydrogen in the terrestrial environment. These findings obliged us to
introduce ambigravitons, i.e. a symmetric graviton and antigraviton
environment. The interaction between antimatter and antimatter then
obeyed the same rules as matter and matter.

If we assume ambigravitons in the early Universe, then it is more
or less self-evident that the matter-antimatter distribution should
also be symmetric as proposed by many authors mentioned in
Section~\ref{s:introd}

\subsection{Antimatter in the gravitational field of the Earth}
\label{ss:3_2}
The most important result expected from the antimatter experiments is, of
course, the behaviour of antihydrogen in the gravitational field of the
Earth. The result determines whether or not a violation of WEP is involved.
This question is very controversially discussed in the literature as
presented in Section~\ref{s:introd}. Our model shown in
Figures~\ref{fig:M_g_am} and \ref{fig:M_ag_am}
leads to the surprising result that there is NO
attraction or repulsion between matter and antimatter in a \emph{symmetric}
antigraviton environment. However, the very small asymmetry mentioned in
Subsections~\ref{ss:2_4} and \ref{ss:2_5} could change the situation.

In view of the results of \citet{Andetal23} presented at the end of
Subsection~\ref{ss:2_4}, it is indeed necessary to reconsider our model
assumptions in detail.
To simplify the calculations, we consider only three attractions with
accelerations of $0.75~g$, $1.04~g$,
and $0.44~g$ for antimatter in the gravitational field of the Earth
(compatible with the best fit of \citet{Andetal23} and the extreme
uncertainty ranges). Can they be understood by our model?

We obtained a zero momentum balance in Figure~\ref{fig:M_g_am}
under perfectly symmetric antigraviton conditions. Several indications have
been mentioned in Section~\ref{s:Results} that a perfect symmetry cannot be
expected. The deviation required can now be calculated with the help of the
observed attractions.

According to Equation~(\ref{eqn:matter_matter}), applied
to the $n$ reduced gravitons in one second, we get a
momentum balance of $+ 2\,n\,Y\,\vec{p}_{\rm G}$ for an attraction of $1\,g$
without or a symmetric antigraviton environment,
see Figure~\ref{fig:M_g_m}.
When Mass~1 converts a small fraction~$x\,Y$ of the negative antigravitons into
negative reduced gravitons, then the balance is
%
\begin{eqnarray}
+ 2\,n\,Y\,\vec{p}_{\rm G} - 2\,n\,\vec{p}_{\rm G}(1 - 2\,Y) \times x\,Y =
\nonumber \\
+ 2\,n\,Y\,\vec{p}_{\rm G} (1 - x + 2\,x\,Y)~~.
\label{eqn:ratio}
\end{eqnarray}

If one antigraviton in Figure~\ref{fig:M_ag_am} (Input~B)
is replaced by a reduced graviton (Input~A),
we get $+ 2\,\bar{\vec{p}}_{\rm G}(1 - Y)$.
The question now is what fraction of the $n$~negative antigravitons
in $1~\s$ must be replaced in
Figure~\ref{fig:M_g_am} to get accelerations of
$+ 0.75\,g, + 1.04\,g, {\rm or} \\
{+ 0.44\,g}$, respectively.
The answer follows from equating
$+ 2\,n\,\bar{\vec{p}}_{\rm G}(1 - Y) \times x\,Y$
with the result of
Equation~(\ref{eqn:ratio}), which leads to a gravitational attraction of
$1\,g$, multiplied by the corresponding conversion fraction:
%
\begin{eqnarray}
+ 2\,n\,\bar{\vec{p}}_{\rm G}(x\,Y - x\,Y^2) = \nonumber \\
+ 2\,n\,\vec{p}_{\rm G} (Y - x\,Y + 2\,x\,Y^2) \times [0.75,~1.04,~0.44] ~~.
\label{eqn:measurements}
\end{eqnarray}
The quadratic terms of $Y$ can be neglected, because of the very small value
of $Y$ (cf. Subsection~\ref{ss:2_3}). The evaluation then gives values of
$x\,Y = [0.43,\,0.51,\,0.31]\,Y$.
They can be adjusted to comply with more precise measurements in the future.
The fraction is, in any case, extremely small and supports the speculations in
Subsection~\ref{ss:2_5} that only very few antigravitons would be converted by
matter.

\subsection{Interactions between particles at close range}   
\label{ss:3_3}

The impact model thus predicts normal gravitational attraction between
remote masses of
matter and matter as well as of antimatter and antimatter, whereas
there is no interaction between matter and antimatter in
a symmetric ambigraviton environment.
In Subsection~\ref{ss:3_2}, processes have been discussed,
if this condition is not met.

As pointed out by \citet{Fle17}, cf. Subsection~\ref{ss:2_5}, this behaviour
changes at close range, where we find a very strong repulsion between
matter and matter as well as between antimatter and antimatter, whereas a
very strong attraction is characteristic for near matter-antimatter
encounters.\footnote{Electrostatic forces are not taken into account.
Neutron and antineutron interactions could be appropriate models.}

\subsection{The cosmological constant problem}   
\label{ss:3_4}

The gravitation between particles and photons cannot be treated without
considering the cosmological constant and the problem of a consistent
interpretation within the standard model. The cosmological constant
has to be many orders of magnitude smaller than predicted by the standard
model. From the immense literature on this topic
values of $10^{-122}$ to $10^{-46}$ can be
quoted \cite[cf., e.g.][]{Abb88,Wei89,Mar12,Plaetal}. In line with our
proposal, we find a suggestion by \citet{Lom19} very interesting
that not every vacuum energy is gravitating.

The gravitational impact model in \citet{WilWilDwi} had assumed
in Equation~16 an emission
coefficient $\eta_{\rm G} = c_0^2/(2 h)$ of quadrupoles by matter
(corresponding to half
the intrinsic de Broglie frequency per kilogram) with $c_0$ the speed of light and $h$
the Planck constant. The justification for assuming half the frequency was
that two emissions of virtual gravitons are involved per graviton emission.
Do we have to adjust the emission coefficient in a matter-antimatter system
with ambigravitons? The answer is no, because the direct interactions
(see Path~1a in Figure~\ref{fig:M_g_m}) and the indirect interactions
(see Path~3a in Figure~\ref{fig:AM_g_am}) equally contribute to the
production of reduced gravitons in the case of matter particles and
reduced antigravitons for antimatter particles. Thus the proposed interaction
between matter and matter particles will not change in an ambigraviton
environment and, for reasons of symmetry, we can expect the same between
antimatter and antimatter particles.

Figure~3 of \citet{WilDwi20} is therefore relevant also under an ambigraviton
environment, and we can obtain some estimates from it. With a
value of the electron radius
$r_{G,e} = (2.43 \pm 0.39) \times 10^{-16}\m$
(smaller than or equal to the classical electron radius)
and an electron mass $m_{\rm e}$, the ambigraviton density
is $\rho_{\rm G} = 2.75 \times 10^{41}\m^{-3}$ and with the
surface mass density
$\sigma_{\rm G} = m_{\rm e}/(4 \pi r_{G,e}^2)$ in Equation~23 of
\citet{WilDwi20}, the
Equation~28 can be written as
%
\begin{eqnarray}
\epsilon_{\rm G} =
T_{\rm G} \rho_{\rm G} = 2 \pi G_{\rm N} \sigma_{\rm G}^2/Y ,
\label{eqn:energy}
\end{eqnarray}
where $\epsilon_{\rm G}$ is the undisturbed ambigraviton energy density,
$T_{\rm G}$ the graviton or antigraviton energy and $G_{\rm N}$ Newton's
constant of gravity. The choice of the reduction parameter $Y$ then determines
the remaining quantities. With a value of $Y = 10^{-20}$, we get
\newline
$\epsilon_{\rm G} = 6.22 \times 10^{10}\J\m^{-3}$ and
$T_{\rm G} = 2.26 \times 10^{-31}\J$.

If we now consider that only the reduced portion of the ambigravitons mediates
the gravitational
force in the impact model, it can be concluded that the relevant
gravitational energy density is \\
$Y~\epsilon_{\rm G} = 6.22 \times 10^{-10}\J\m^{-3}$, corresponding to
estimates of the dark energy in the literature,
cf., e.g. \citet{BecMac,Frie08}.

\section{Conclusion}  
\label{s:Concl}
From our hypothetical impact model of gravity
applied to matter-antimatter
configurations in Figures~\ref{fig:M_g_m} to \ref{fig:M_ag_am}, we
expect neither attraction nor repulsion between matter and antimatter
bodies at large distances and strong attraction at short ranges.
The gravitational forces between antimatter and antimatter particles
should be the same as in matter-matter systems. These predictions
are based on a symmetric antigraviton environment. Recent measurements
published in \citet{Andetal23} indicate that we must not assume a perfect
symmetry of antigravitons near matter particles.
The measurements could be explained by a minor
adjustment of the model assumptions concerning the fraction of
indirect to direct interactions.

\section*{Acknowledgements}
This research has made extensive use of the Astrophysics Data System (ADS).
Administrative support has been provided by the Max-Planck-Institute for
Solar System Research, G\"ottingen, Germany.
We thank two Reviewers for many constructive
and valuable comments which improved the presentation of the manuscript.

\section*{Data Availability}  
The physical quantities discussed in Section~\ref{ss:3_4} are taken from
\citet{WilDwi20};\\ DOI:10.5772/intechopen.86744.


\begin{thebibliography}{00}

%
\bibitem[\protect\citeauthoryear{Abbott}{1988}]{Abb88}
Abbott, L. 1988,
The mystery of the cosmological constant,
Scientific American, 258:5,  106--113
%
\bibitem[\protect\citeauthoryear{Ahmadi et al.}{2017}]{Ahmetal17}
Ahmadi, M., Alves, B.\,X.\,R., Baker, C.\,J. and 51~co-authors, 2017,
Observation of the 1S - 2S transition in trapped antihydrogen.
\nat, 541, 506--510
%
\bibitem[\protect\citeauthoryear{Ahmadi et al.}{2018a}]{Ahmetal18a}
Ahmadi, M., Alves, B.\,X.\,R., Baker, C.\,J. and 46~co-authors, 2018a,
Characterization of the 1S-2S transition in antihydrogen,
\nat, 557, 71--75
%
\bibitem[\protect\citeauthoryear{Ahmadi et al.}{2018b}]{Ahmetal18b}
Ahmadi, M., Alves, B.\,X.\,R., Baker, C.\,J., Bertsche W. and \\
48~co-authors , 2018b,
Observation of the 1S--2P Lyman-$\alpha$ transition in antihydrogen,
\nat, 561, 211--215
%
\bibitem[\protect\citeauthoryear{Ahmadi et al.}{2020}]{Ahmetal20}
Ahmadi, M., Alves, B.\,X.\,R., Baker, C.\,J., Bertsche W. and \\
48~co-authors (The ALPHA Collaboration), 2020,
Investigation of the fine structure of antihydrogen,
\nat, 578, 375--379 (cf. Author correction, 2021, \nat, 594, E5)
%
\bibitem[\protect\citeauthoryear{Alfv\'{e}n}{1965}]{Alf65}
Alfv\'{e}n, H., 1965,
Antimatter and the development of the Megagalaxy,
Rev. Mod. Phys., 37, No. 4, 652--665
%
\bibitem[\protect\citeauthoryear{Alfv\'{e}n \& Klein}{1963}]{AlfKle}
Alfv\'{e}n, H., Klein, O., 1963,
Matter-antimatter annihilation and cosmology,
Arkiv f\"or Fysik, 23, 187--194
%
\bibitem[\protect\citeauthoryear{Amole et al.}{2012}]{Amoetal12}
Amole, C., Ashkezari, M.\,D., Baquero-Ruiz, M., Bertsche, W.
and 39~co-authors, 2012,
Resonant quantum transitions in trapped antihydrogen atoms,\\
\nat, 483, Issue 7390, 439--443
%
\bibitem[\protect\citeauthoryear{Amole et al.}{2014}]{Amoetal14}
Amole, C., Andresen, G.\,B., Ashkezari, M.\,D. and 53~co-authors, 2014,
The ALPHA antihydrogen trapping apparatus,
Nuclear Inst. Meth. Phys. Res., A, 735, 319--340
%
\bibitem[\protect\citeauthoryear{Amsler et al.}{2021}]{Amsetal21}
Amsler, C., Antonello, M., Belov, A., Bonomi, G. and 52~co-authors, 2021,
Pulsed production of antihydrogen,
Communications Physics, 4:19, 1--11
%
\bibitem[\protect\citeauthoryear{Anderson}{1932}]{And32}
Anderson, C.\,D., 1932,
The apparent existence of easily deflectable positives,
Science, 76, 238--239
%
\bibitem[\protect\citeauthoryear{Anderson}{1933}]{And33}
Anderson, C.\,D., 1933,
The positive electron,\\
Physical Review, 43, 491--494
%
\bibitem[\protect\citeauthoryear{Anderson et al.}{2023}]{Andetal23}
Anderson, E.\,K., Baker, C.\,J., Bertsche, W. and 68~co-authors, 2023,
Observation of the effect of gravity on the motion of antimatter.
\nat, 621, 716--722
%
\bibitem[\protect\citeauthoryear{Anderson}{1934}]{And34}
Anderson, C.\,D., 1934,
The positron,
\nat, 133, 313--316
%
\bibitem[\protect\citeauthoryear{Banik \& Kroupa}{2020}]{BanKro}
Banik, I., Kroupa, P., 2020,
Solar System limits on gravitational dipoles,
\mnras, 495, 3974--3980
%
\bibitem[\protect\citeauthoryear{Beck \& Mackey}{2007}]{BecMac}
Beck, C., Mackey, M.\,C., 2007,
Measurability of vacuum fluctuations and dark energy.\\
Physica A, 2007, 379, 101--110
%
\bibitem[\protect\citeauthoryear{Bertsche}{2018}]{Ber18}
Bertsche, W.\,A., 2018,
Prospects for comparison of matter and antimatter gravitation with ALPHA-g,\\
Phil. Trans. R. Soc. A 2018, 376, 20170265
%
\bibitem[\protect\citeauthoryear{Benoit-Levy \& Chardin}{2012}]{BenCha}
Benoit-Levy, A., Chardin, G., 2012,
Introducing the Dirac-Milne Universe,
\aap, 537, A78, 1--12
%
\bibitem[\protect\citeauthoryear{Borchert et al.}{2022}]{Bor22}
Borchert, M.\,J., Devlin1, J.\,A., Erlewein, S.\,R., Fleck, M. and
19~co-authors, 2022,
A 16-parts-per-trillion measurement of the antiproton-to-proton
charge - mass ratio,\\
\nat, 601, 53--57
%
\bibitem[\protect\citeauthoryear{Brusa et al.}{2017}]{Bruetal17}
Brusa, R.\,S., Amsler, C., Ariga, T., Bonomi, G. and 60~co-authors, 2017,
The AEgIS experiment at CERN: measuring antihydrogen free-fall in Earth's
gravitational field to test WEP with antimatter,\\
J. Phys.: Conf. Ser., 791, 012014, 1--8
%
\bibitem[\protect\citeauthoryear{Cabbolet}{2010}]{Cab10}
Cabbolet, M.\,J.\,T.\,F., 2010,
Elementary Process Theory: a formal axiomatic system with a potential
application as a foundational framework for physics supporting gravitational
repulsion of matter and antimatter,
Ann. Phys. (Berlin) 522:10, 699--738 (cf., Corrigendum,\\
Ann. Phys. (Berlin) 528:7--8, 626--627, 2016)
%
\bibitem[\protect\citeauthoryear{Caldwell \& Dvali}{2020}]{CalDva}
Caldwell, A., Dvali, G., 2020,
On the gravitational force on anti-matter,
Fortschr. Phys., 2000092, 1--2
%
\bibitem[\protect\citeauthoryear{Chardin \& Manfredi}{2018}]{ChaMan}
Chardin, G., Manfredi, G., 2018,
Gravity, antimatter and the Dirac-Milne universe,\\
Hyperfine Interactions, 239, id.45
%
\bibitem[\protect\citeauthoryear{Chardin et al.}{2021}]{Chaetal21}
Chardin, G., Yohan Dubois, Y., Manfredi G., Miller, B., Stah, C., 2021,
MOND-like behavior in the Dirac--Milne universe.
Flat rotation curves and mass versus velocity relations in galaxies
and clusters,
\aap, 652, A91
%
\bibitem[\protect\citeauthoryear{Crivelli \& Kolachevsky}{2020}]{CriKol}
Crivelli, P., Kolachevsky, N., 2020,
Optical trapping of antihydrogen towards an atomic anti-clock,
Hyperfine Interactions, 241, Issue 1, article id. 60
%
\bibitem[\protect\citeauthoryear{Dirac}{1928}]{Dir28a}
Dirac, P.\,A.\,M., 1928,
The quantum theory of the electron,
Proc. Royal Soc. London, Series A, 117,\\ Issue 778, 610--624
%
\bibitem[\protect\citeauthoryear{Dine \& Kusenko}{2004}]{DinKus}
Dine, M., Kusenko, A., 2004,
Origin of the matter-antimatter asymmetry,
Rev. Mod. Phys., 76, 1--30
%
\bibitem[\protect\citeauthoryear{Drude}{1897}]{Dru97}
Drude, P., 1897,
Ueber Fernewirkungen.\\
Verlag Barth (Leipzig), 1897
%
\bibitem[\protect\citeauthoryear{Du}{2023}]{Du2023}
Du, H., 2023,
ALPHA-g experiment gives experimental support for repulsive antigravity.
Preprint, December 2023,\\ Doi: 10.13140/Rg.2.2.19690.70086
%
\bibitem[\protect\citeauthoryear{Einstein}{1911}]{Ein11}
Einstein, A., 1911,
\"Uber den Einflu{\ss} der Schwerkraft auf die Ausbreitung des Lichtes,\\
Ann. Phys. (Leipzig), 340, 898--908
%
\bibitem[\protect\citeauthoryear{Einstein}{1916}]{Ein16}
Einstein, A., 1916,
Die Grundlage der allgemeinen Relativit\"atstheorie,\\
Ann. Phys. (Leipzig), 354, Issue 7, 769--822
%
\bibitem[\protect\citeauthoryear{E\"otv\"os, Pek\'ar \& Fekete}{1922}]{Eoeetal}
v. E\"otv\"os, R., Pek\'ar, D., Fekete, E., 1922,
Beitr\"age zum Gesetze der Proportionalit\"at von Tr\"agheit und Gravit\"at,
Ann. Phys. (Leipzig), 373, 11--66
%
\bibitem[\protect\citeauthoryear{Fatio de Duilleir}{1690}]{Fat90}
Fatio de Duilleir, N., 1690,
De la cause de la pesanteur,\\
Not. Rec. Roy. Soc. London, 6, 2, 125
%
\bibitem[\protect\citeauthoryear{Fleming}{2017}]{Fle17}
Fleming, R., 2017,
The matter-antimatter dipole,\\
The General Science Journal, October 1, 1--7
%
\bibitem[\protect\citeauthoryear{Frieman, Turner \& Huterer}{2008}]{Frie08}
Frieman, J.\,A., Turner, M.\,S., Huterer, D., 2008,
Dark energy and the accelerating universe,\\
Annu. Rev. Astron. Astrophys., 46, 385--432
%
\bibitem[\protect\citeauthoryear{Good}{1961}]{Goo61}
Good, M.\,L., 1961,
$K_2^0$ and the Equivalence Principle,\\
Physical Review, 121, 311--314
%
\bibitem[\protect\citeauthoryear{Goldhaber}{1956}]{Gol56}
Goldhaber, M., 1956,
Speculations on cosmogony,\\
Science, 124, Issue 3214, 218--219
%
\bibitem[\protect\citeauthoryear{Hajdukovic}{2010}]{Haj10}
Hajdukovic, D.\,S., 2010,
On the vacuum fluctuations, Pioneer Anomaly and Modified Newtonian
Dynamics,\\
\apss, 330, 207--209
%
\bibitem[\protect\citeauthoryear{Hajdukovic}{2013}]{Haj13}
Hajdukovic, D.\,S., 2013,
Can observations inside the Solar System reveal the gravitational
properties of the quantum vacuum?
\apss, 343, 505--509
%
\bibitem[\protect\citeauthoryear{Hajdukovic}{2020a}]{Haj20a}
Hajdukovic, D.\,S., 2020a,
Antimatter gravity and the Universe,
Mod. Phys. L. A, 35, 2030001
%
\bibitem[\protect\citeauthoryear{Hajdukovic}{2020b}]{Haj20b}
Hajdukovic, D.\,S., 2020b,
On the gravitational field of a point-like body immersed in a quantum
vacuum,\\
\mnras, 491, 4816--4828
%
\bibitem[\protect\citeauthoryear{Hajdukovic \& Walter}{2021}]{HajWal}
Hajdukovic, D.\,S., Walter, S., 2021,
Gravitational polarization of the quantum vacuum caused by
two point-like bodies,
\mnras, 503, 5091--5099
%
\bibitem[\protect\citeauthoryear{Iorio}{2019}]{Ior19}
Iorio, L., 2019,
A comment on ``Can observations inside the Solar System reveal the
gravitational properties of the quantum vacuum?'' by D.S. Hajdukovic,\\
\apss, 364, Issue 8, article id. 126, 2 pp.
%
\bibitem[\protect\citeauthoryear{Kellerbauer et al.}{2012}]{Keletal}
Kellerbauer, A., Allkofer, Y., Amsler, C., Belov, A.\,S. and 62~co-authors,
2012,
The AEgIS experiment at CERN.
Measuring the free fall of antihydrogen,\\
Hyperfine Interactions, 209,  43--49
%
\bibitem[\protect\citeauthoryear{Kleinknecht}{2001}]{Klein01}
Kleinknecht, K., 2001,
Violation of matter-antimatter symmetry,
Ann. Phys. (Berlin) 513:1, 133--150
%
\bibitem[\protect\citeauthoryear{Lombriser}{2019}]{Lom19}
Lombriser, L., 2019,
On the cosmological constant problem,
Physics Letters B, 797, 134804
%
\bibitem[\protect\citeauthoryear{L\"uders}{1957}]{Lue57}
L\"uders, G., 1957,
Proof of the TCP theorem,\\
Annals of Physics, 2, 1--15
%
\bibitem[\protect\citeauthoryear{Martin}{2012}]{Mar12}
Martin, J., 2012,
Everything you always wanted to know about the cosmological constant problem
(but were afraid to ask),
Comptes Rendus Physique, 13, 566--665
%
\bibitem[\protect\citeauthoryear{Mansouli\'{e}}{2019}]{Man19}
Mansouli\'{e}, B., 2019,
Status of the GBAR experiment at CERN,
Hyperfine Interactions, 240, Issue 1, 1--6
%
\bibitem[\protect\citeauthoryear{Morrison}{1958}]{Mor58}
Morrison, P., 1958,
Approximate nature of physical symmetries,
Am. J. Phys., 26, 358--368
%
\bibitem[\protect\citeauthoryear{Nieto \& Goldman}{1991}]{NieGol}
Nieto, M.\,M., Goldman, J.\,T., 1991,
The arguments against ``antigravity`'' and the gravitational acceleration of
antimatter,
\physrep, 205, 221--281
%
\bibitem[\protect\citeauthoryear{Ogievetsky \& Polubarinov}{1965}]{OgiPol}
Ogievetsky, V.\,I., Polubarinov, I.\,V., 1965,
Interacting field of spin 2 and the Einstein equations,\\
Ann. Phys., 36, 167--208
%
\bibitem[\protect\citeauthoryear{Okun}{2000}]{Oku00}
Okun, L.\,B., 2000,
Photons and static gravity,\\
Mod. Phys. Lett. A, 15, 1941--1947
%
\bibitem[\protect\citeauthoryear{Okun, Selivanov \& Telegdi}{2000}]{Okuetal}
Okun, L.\,B., Selivanov, K.\,G., Telegdi, V.\,L., 2000,
On the interpretation of the redshift in a static gravitational field,
Am. J. Phys., 68, 115--119
%
\bibitem[\protect\citeauthoryear{Parthey et al.}{2011}]{Paretal11}
Parthey, C. G., Matveev, A., Alnis, J., Bernhardt, B., and 13~co-authors,
2011,
Improved measurement of the hydrogen 1S - 2S transition frequency,\\
\prl, 107, 203001-1--5
%
\bibitem[\protect\citeauthoryear{P\'{e}rez et al.}{2015}]{Peretal15}
P\'{e}rez, P., Banerjee, D., Biraben, F., Brook-Roberge, D. and
57~co-authors, 2015,
The GBAR antimatter gravity experiment,
Hyperfine Interact, 233, 21--27
%
\bibitem[\protect\citeauthoryear{Planck Collaboration}{2020}]{Plaetal}
Planck Collaboration: 181 co-authors, 2020,
Planck 2018 results VI. Cosmological parameters,\\
\aap, 641, A6, 1--67
%
\bibitem[\protect\citeauthoryear{Rovelli}{2000}]{Rov00}
Rovelli, C., 2000,
Notes for a brief history of quantum gravity,
arXiv:gr-qc/0006061
%
\bibitem[\protect\citeauthoryear{Sakharov}{1966}]{Sak66}
Sakharov, A.\,D., 1966,
Violation of CP invariance, C asymmetry, and baryon asymmetry of the
universe,\\
Soviet Physics JETP, 22, No. 1, 241--249
%
\bibitem[\protect\citeauthoryear{Scherk}{1979}]{Sch79}
Scherk, J., 1979,
Antigravity: a crazy idea?\\
Phys. Lett., 88B,  265--267
%
\bibitem[\protect\citeauthoryear{Schiff}{1959}]{Sch59}
Schiff, L.\,I., 1959,
Gravitational properties of antimatter,
Proc. Nat. Acad. Sci., 45, 69--80
%
\bibitem[\protect\citeauthoryear{Schuster}{1898}]{Sch98}
Schuster, A., 1898,
Potential matter--A holiday dream,
\nat, 58, 367
%
\bibitem[\protect\citeauthoryear{Testera et al.}{2015}]{Tesetal15}
Testera, G., Aghion, S., Amsler, C., Ariga, T. and 70~co-authors, 2015,
The AEgIS experiment,\\
Hyperfine Interactions, 233, 13--20
%
\bibitem[\protect\citeauthoryear{Touboul et al.}{2019}]{Touetal19}
Touboul, P., M\'{e}tris, G., Rodrigues, M., Andr\'{e}, Y. and 39~co-authors,
2019,
Space test of the equivalence principle: first results of the
MICROSCOPE mission,\\
Class. Quantum Grav., 36, 225006, 1--34
%
\bibitem[\protect\citeauthoryear{Villata}{2011}]{Vil11}
Villata, M., 2011,
CPT symmetry and antimatter gravity in general relativity,
Europhys. L., 94, 20001, 1--4
%
\bibitem[\protect\citeauthoryear{Villata}{2013}]{Vil13}
Villata, M., 2013,
On the nature of dark energy: the lattice Universe,
\apss, 345, 1--9
%
\bibitem[\protect\citeauthoryear{Villata}{2015}]{Vil15}
Villata, M., 2015,
The matter-antimatter interpretation of Kerr spacetime,\\
Ann. Phys. (Berlin) 527, No. 7--8, 507--512
%
\bibitem[\protect\citeauthoryear{Vlasov}{1956}]{Vla65}
Vlasov, N.\,A., 1965,
Optical search for antimatter in the Universe,
Soviet Astronomy--AJ, 8, No. 5, 715--718
%
\bibitem[\protect\citeauthoryear{Weinberg}{1989}]{Wei89}
Weinberg, S., 1989,
The cosmological constant problem,
Rev. Mod. Phys., 61, 1--23
%
\bibitem[\protect\citeauthoryear{Wilhelm \& Dwivedi}{2014}]{WilDwi14}
Wilhelm, K., Dwivedi, B.\,N., 2014,
Secular perihelion advances of the inner planets and Asteroid Icarus,\\
\na, 31, 51--55
%
\bibitem[\protect\citeauthoryear{Wilhelm \& Dwivedi}{2015}]{WilDwi15}
Wilhelm, K., Dwivedi, B.\,N., 2015,
On the potential energy in a gravitationally bound two-body system
with arbitrary mass distribution,
arXiv 1502.05662
%
\bibitem[\protect\citeauthoryear{Wilhelm \& Dwivedi}{2018a}]{WiDwGala}
Wilhelm, K., Dwivedi, B.\,N., 2018a,
A physical process of the radial acceleration of disc galaxies,\\
\mnras, 474, Issue 4, 4723--4729
%
\bibitem[\protect\citeauthoryear{Wilhelm \& Dwivedi}{2018b}]{WiDwGalb}
Wilhelm, K., Dwivedi, B.\,N., 2018b,
On the radial acceleration of disc galaxies,\\
\mnras, 494, Issue 3, 4015--4025
%
\bibitem[\protect\citeauthoryear{Wilhelm \& Dwivedi}{2019}]{WilDwi19}
Wilhelm, K., Dwivedi, B.\,N., 2019,
Gravitational redshift and the vacuum index of refraction,\\
\apss, 364, Issue 2, article id. 26, 7 pp
%
\bibitem[\protect\citeauthoryear{Wilhelm \& Dwivedi}{2020}]{WilDwi20}
Wilhelm, K., Dwivedi, B.\,N., 2020,
Impact models of gravitational and electrostatic forces (Ref.~2),\\
Planetology (ed.~B. Palaszewski), Chapter 5, 1--39
%
\bibitem[\protect\citeauthoryear
{Wilhelm, Wilhelm \& Dwivedi}{2013}]{WilWilDwi}
Wilhelm, K., Wilhelm, H., Dwivedi, B.\,N., 2013,
An impact model of Newton's law of gravitation (Ref.~1)\\
\apss, 343, 135
%
\bibitem[\protect\citeauthoryear{Willmann \& Jungmann}{2016}]{WillJung}
Willmann, L., Jungmann, K., 2016,
Matter-antimatter asymmetry\,--\,aspects at low energy,\\
Ann. Phys. (Berlin) 528:1--2, 108--114
%
\end{thebibliography}
\end{document}